\documentclass[slac_one]{revtex4}

\usepackage{graphicx}

\usepackage{fancyhdr}

\usepackage{wrapfig}
\usepackage{units}

\newcommand{\pslash}{p\llap{/\kern+0.1em}}
\newcommand{\GeV}{\mathrm{G\kern-0.01eme\kern-0.11emV}}
\newcommand{\MeV}{M\kern-0.02eme\kern-0.11emV}
\newcommand{\keV}{ke\kern-0.11emV}
\newcommand{\eV}{e\kern-0.11emV}

\begin{document}


\title{Heavy quark production and spectroscopy at HERA} %% Paper title goes here


%




\author{M. J\"ungst\,\,(on behalf of the H1 and ZEUS collaborations)}

\affiliation{Physikalisches\phantom{0}Institut\phantom{0}Universit\"at\phantom{0}Bonn\\
Nu{\ss}allee\phantom{0}12,\phantom{0}53115\phantom{0}Bonn,\phantom{0}Germany}

%

%\author{P. Lucas}

%\affiliation{FNAL, Batavia, IL 60510, USA}

\begin{abstract}

Heavy flavour production and spectroscopy are key components of the HERA
physics programme. I will summarise a selection of the recent
results obtained by the H1 and ZEUS collaborations. The production of excited
charm mesons and $J/\psi$ will be discussed as well as measurements of b quark
cross-sections in photoproduction. The status of searches for exotic bound
states and the $D^{*}p$ resonance will be updated.
\end{abstract}


\maketitle

\thispagestyle{fancy}

\vspace{0.2cm}
\section{INTRODUCTION}
Heavy flavour production in $e^{\pm}p$ collisions at HERA provides a good testing
ground of perturbative Quantum Chromodynamics (pQCD) as the high quark mass
provides a hard scale. Furthermore, other hard scales such as $Q^2$, the virtuality of
the exchanged boson, or $p_t$, the transverse momentum of the heavy
quark, allow resummation techniques to be tested.
Measurements of production rate and kinematic
properties of heavy quark bound states such as charmonium also give direct access
to the non-perturbative part of the production process.
New theoretical models can also be tested by searches for exotic
bound states like the $D^{*}p$ resonance.\par
Different kinematic variables are used to describe the $ep$ interaction
at HERA: $Q^2$, the Bjorken scaling variable, $x$, and the inelasticity, $y$.
Until 1997 HERA ran at a centre-of-mass energy of
$\sqrt{s}=300\,\text{GeV}$. This energy was increased to
$\sqrt{s}=320\,\text{GeV}$ for data taken from 1998 onwards.
Due to improvements of the detector and accelerator during a break in the data
taking, the available dataset is split into two periods. In the HERA\,I period
from 1996 until 2000 about $130\,\text{pb}^{-1}$ and between 2003
and 2007 of HERA\,II about $400\,\text{pb}^{-1}$ per experiment were collected.

The kinematic range of the analysed data can be separated in the following
two regimes: photoproduction ($\gamma p$), where the exchanged photon in the
process is almost real, and deep inelastic scattering (DIS), where the
exchanged photon is virtual. Experimentally,
$\gamma p$ is defined by the scattered electron not being in
the acceptance region of the detector, corresponding to a cut $Q^2 \lesssim 1\,\text{GeV}^2$.
%Achim:
%Zeile 10: About 150 pb-1 were collected per experiment in the
%period from 1996 to 2000, and about 500 pb-1 from 2003 to 2007.
%(Das sind die Rohzahlen. Fuer Physik war's weniger. Korrigieren?)

\section{THEORY}
There are different approaches for the calculation of heavy
flavour production in next-to-leading order perturbative QCD. The
massive approach assumes no initial charm or beauty in the
proton (or photon). Heavy flavours are only generated dynamically from
the gluon distribution. This approach is particularly valid if the heavy quark
mass is of the order of other hard scales like $Q^2$ or the transverse
momentum $p_t$ of the heavy quarks.
In the massless approach, the leading-logarithmic and next-to-leading
logarithmic terms for example in $\alpha_s\log Q^2/m^2_{c,b}$
are resummed. This approach assumes the heavy quark to be massless and
is therefore only valid if other quantities like $Q^2$ or $p_t$
provide the dominant scale. Models using the  $k_{T}$
factorisation approach~\cite{kt} are based on non-collinear parton dynamics
based on the CCFM~\cite{cascade} evolution equations.
In heavy quark bound states like charmonium, only the production of the
$c\bar{c}$ pair can be described in pQCD, whereas the formation of the $J/\psi$
bound state which occurs at long distances has to be described by
phenomenological models. In the framework of non-relativistic QCD (NRQCD)~\cite{nrqcd}
so-called colour singlet (CS) and colour octet (CO) states coexist.
In the spectroscopy of heavy quark bound states containing only one heavy
quark, Heavy Quark Effective Theory (HQET) is used to predict the
production and decay properties.

\section{BEAUTY IN PHOTOPRODUCTION}

\begin{wrapfigure}[10]{r}{0.38\textwidth}
  \vspace{-35pt}
  \begin{center}
    \includegraphics[bb = 30 0 510 370,clip=true,width=60mm]{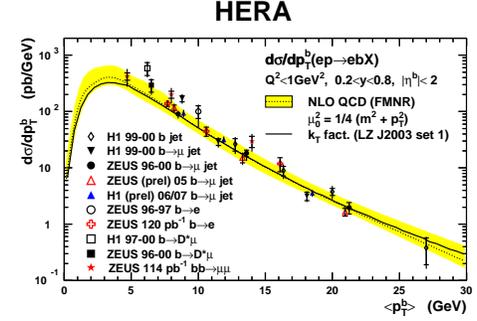}
\end{center}
  \vspace{-20pt}
\caption{Cross sections for beauty production as a function of $p^{b}_{T}$
from various decay channels.}
\label{beauty}
\end{wrapfigure}
An important test of pQCD is provided by open heavy flavour production. In Figure
\ref{beauty} the measurements of the beauty cross section in photoproduction
of both the H1 and the ZEUS collaborations are shown as a function of the $b$
quark transverse momentum, $p_{T}$. The analyses used different datasets, techniques and decay channels
providing independent measurements cross-checking each other.
The different measurements agree well with each other and are in reasonable
agreement with the NLO prediction from FMNR\cite{FMNR}.
They cover a wide range in $p_{T}^{b}$ giving a consistent picture of $b$
quark photoproduction.

\vspace{0.5cm}

\section{INELASTIC $\mathbf{J/\psi}$ PRODUCTION}
At HERA, the charmonium state $J/\psi$ is produced predominantly by the boson gluon fusion (BGF) process.
The theoretical models which are used to compare with the measurements follow
either the DGLAP evolution or use the $k_{T}$ factorisation approach.
After the heavy quark pair ($c\bar{c}$) is produced at short distances,
the formation of the $J/\psi$ bound state is described by non-perturbative
long distance matrix elements (LDME). In the NRQCD models, contributions
from both CS and CO states are predicted; one aim of the measurements is the
extraction of their relative contributions.
\begin{wrapfigure}[13]{r}{0.38\textwidth}
  \begin{center}
  \vspace{-20pt}
\includegraphics[bb = 0 0 520 370, clip=true,width=0.36\textwidth]{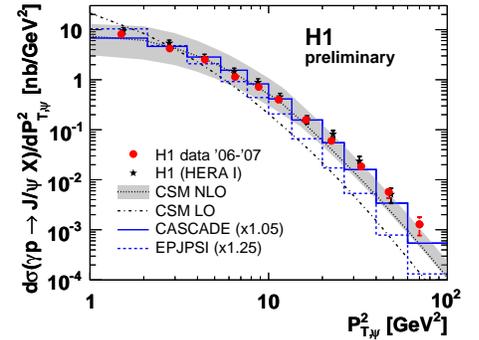}
\end{center}
  \vspace{-20pt}
\caption{Differential cross section as a function of $p_{T,\psi}^{2}$ in the
  ($\gamma p$) regime}
\label{jpsi}
\end{wrapfigure}
The H1 collaboration measured inelastic $J/\psi$ production
with decays to $\mu^+\mu^-$ in the photoproduction region using $\sim
\unit{166}{\text{pb}^{-1}}$ from the 2006-2007 data, complementing their previous
measurement in the DIS region~\cite{h1jpsi}.
In Figure ~\ref{jpsi} the differential cross section as a function of
$p_{T}^{2}$ is shown and compared with Monte Carlo predictions and the CS
calculations at leading order (LO) and at next-to-leading order (NLO).
The scaled CASCADE~\cite{cascade} Monte Carlo, which uses the $k_{T}$ factorisation in the
colour changing flavour mode, is able to reproduce the slope better than the
scaled EPJPSI~\cite{epjpsi} Monte Carlo, following the DGLAP evolution.
While the CS LO calculation is not able to describe the data, there is a good
agreement with the NLO calculation within the large normalisation uncertainties.
As the CS provides a generally good description of the data when using
the $k_{T}$ factorisation or calculations at higher orders, no significant
colour octet contribution is required, although there are still
large normalisation uncertainties on the prediction.

In order to reduce the effect of the normalisation uncertainty,
polarisation measurements are an important testing ground for the NRQCD
predictions. Normalised quantities can be measured using helicity parameters
to characterise the decay angular distributions. A measurement of the $J/\psi$
helicity~\cite{zeusjps} has been performed by the ZEUS collaboration using
the complete HERA statistics ($\sim \unit{470}{\text{pb}^{-1}}$).
The angular distributions of the $J/\psi\rightarrow l^{+}l^{-}$ decay can be
parametrised as $\frac{d^{2}\sigma}{d\Omega d y}\propto 1 +  \lambda(y)\cos^2{\Theta} + \mu(y)\sin{2\Theta}\cos{\phi} +\frac{1}{2}
    \nu(z)\sin{^2\Theta}\cos{^2 \phi}$ ~\cite{beneke}.
%$\frac{d^{2}\sigma}{d\Omega d y}\propto 1 +  \lambda(y)\cos^2{\Theta} + \mu(y)\sin{2\Theta}\cos{\phi} +\frac{1}{2}
%\nu(z)\sin{^2\Theta}\cos{^2 \phi}$ ~\cite{beneke}.\\
The following integrated helicity formulae are used, depending on the
chosen reference frame:\\[0.2cm]
\begin{minipage}{0.06\textwidth}
~\\
\end{minipage}
\begin{minipage}{0.39\textwidth}
  \begin{tabular}{l l}
    $\frac{1}{\sigma}\frac{d^{2}\sigma}{d \cos{\Theta}d z}$ &$\propto 1 +\lambda(z)\cos^2{\Theta}$,\\
    $\frac{1}{\sigma}\frac{d^{2}\sigma}{d \phi d z}$ & $\propto 1 + \frac{\lambda(z)}{3}+\frac{\nu(z)}{3}\cos{^2\phi}$,\\[0.2cm]
  \end{tabular}
\end{minipage}
\par\noindent where $\theta$ is the angle between the $\mu^{+}$ vector in the $J/\psi$ rest
frame and the z axis, and $\phi$ is the azimuthal angle in the x-y plane of
the $\mu^{+}$ vector in the $J/\psi$ rest frame.\\
\begin{wrapfigure}[15]{r}{0.38\textwidth}
  \vspace{-20pt}
  \begin{center}
\includegraphics[bb = 0 0 530 530, clip=true,width=0.33\textwidth]{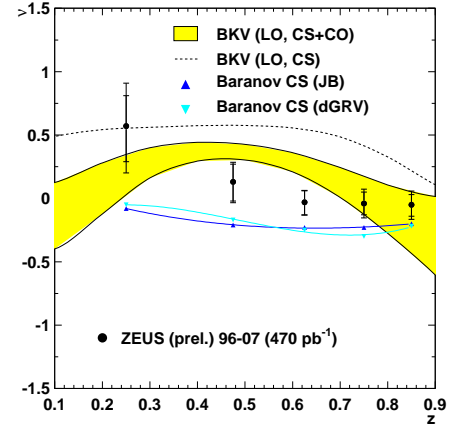}
\end{center}
  \vspace{-20pt}
\caption{Helicity parameter, $\nu$, as a function of the inelasticity, $z$}
\label{jpsihel}
\end{wrapfigure}
Figure~\ref{jpsihel} shows the measurement of the helicity parameter $\nu$ as
a function of the inelasticity $z$.
In addition to the NRQCD predictions two different predictions following the
$k_{t}$ factorisation approach are compared with the measurement.
The LO NRQCD predictions (BKV)~\cite{beneke} do not describe the dependency of this
parameter well, whereas the prediction including the CO contribution seems to be
favoured within the large uncertainties.
The other two predictions (Baranov)~\cite{baranov}, predict a small negative
polarisation and are always below the measurement.
In this scheme where only CS contributions have been taken into account two
different parametrisations of un-integrated gluon distributions have been used.
The CS model with $k_{T}$ factorisation gives predictions that are more similar to
CS+CO compared to the CS only prediction.
To distinguish between the different theoretical contributions the large
variation between the calculations has to be understood and it would be good
if the theoretical predictions could be improved and made available at NLO.
\vspace{-0.3cm}
\section{SPECTROSCOPY}
\vspace{-0.2cm}
 \begin{wrapfigure}[15]{r}{0.38\textwidth}
  \vspace{-30pt}
  \begin{center}
    \includegraphics[bb = 0 0 422 506, clip=true,width=0.36\textwidth]{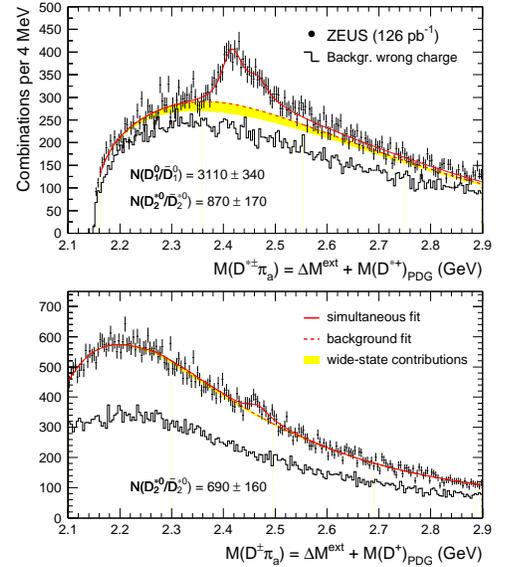}
\end{center}
  \vspace{-20pt}
\caption{$M(D^{*+}\pi_{a})$ and $M(D^{+}\pi_{a})$ distributions for the
  $D_{1}(2420)^{0}$ and the  $D_{2}^{*}(2460)^{0}$ candidates.}
\label{spect}
\end{wrapfigure}
The large charm production cross section at HERA permits measurements to be
made of excited charm and charm-strange mesons.
The production of excited charm and charm-strange mesons is observed by the
ZEUS collaboration using the full HERA\,I dataset corresponding to an
integrated luminosity
of $\sim \unit{126}{\text{pb}^{-1}}$ ~\cite{zeusmeson:08}.
The following decay channels were investigated:\\
\begin{minipage}{0.06\textwidth}
~\\
\end{minipage}
\begin{minipage}{0.39\textwidth}
  \begin{tabular}{l l}
    $D_{1}(2420)^{0}$ & $\rightarrow D^{*\pm}\pi^{\mp}$\\
    $D_{2}^{*}(2460)^{0}$ & $\rightarrow D^{*\pm}\pi^{\mp}$\\
    & $\rightarrow D^{\pm}\pi^{\mp}$\\
    $D_{S1}(2536)^{0}$ & $\rightarrow D^{*+}K^{0}_{s}$\\
    & $\rightarrow D^{*0}K^{+}_{s}$\\
  \end{tabular}
\end{minipage}

\par\noindent Figure~\ref{spect} shows the $M(D^{*+}\pi_{a})$ and $M(D^{+}\pi_{a})$
distributions for the charm meson candidates reconstructed in the given
decay channels.
The measured masses of the observed mesons have been found to be in reasonable
agreement with the world average values.
In addition to the masses also the widths, helicity and the relative branching
fractions could be extracted.
The measured $D_{1}^{0}$ width is $\Gamma(D^{0}_{1})=53.2 \pm 7.2(\text{stat.})^{+3.3}_{-4.9}(\text{syst.})$ MeV
which is above the world average value of $20.4 \pm 1.7$ MeV.
A larger S-wave admixture at ZEUS with respect to that in measurements with
restricted phase space could explain the differences as already a small S-wave
admixture could have sizeable contributions to the $D_{1}^{0}$ width.
The measured value for the helicity parameter of $h(D_{1}^{0}) =
5.9^{+3.0}_{-1.7}(\text{stat.})^{+2.4}_{-1.0}(\text{syst.})$  has to be compared with the
prediction of the HQET for a pure S-wave ($h=0$) and a pure D-wave ($h=3$).
For the charm-strange meson $D_{s1}^{+}$ the measured parameter is $h(D_{s1}^{+}) =
-0.74^{+0.23}_{-0.17}(\text{stat.})^{+0.06}_{-0.05}(\text{syst.})$ which is inconsistent
with the prediction for a pure D-wave and more than two standard deviations
away from the prediction for a pure S-wave.
So the measurement suggests a significant contribution of both D-wave and S-wave
amplitudes to the $D_{s1}(2536)^{+}\rightarrow D^{*+}K_{S}^{0}$ decay.
In addition, a search for the radially excited charm meson
$D^{*'}(2640)^{\pm}$has been made, which was reported by DELPHI as a narrow resonance in the final
state $D^{*\pm}\pi^{+}\pi^{-}$ at 2637 MeV.
In the inspected mass range no signal was observed and according to the
expected mass and width the following limit at ($95 \%$ confidence level) was extracted:
$f(c\rightarrow D^{*'+})\cdot\mathcal{B}_{D^{*'+}\rightarrow D^{*+}\pi^{+}\pi^{-}}<0.4 \% $.

\section{D*p RESONANCE}
 \begin{wrapfigure}[21]{r}{0.4\textwidth}
  \vspace{-30pt}
  \begin{center}
\includegraphics[bb = 0 0 550 390, clip=true,width=0.38\textwidth]{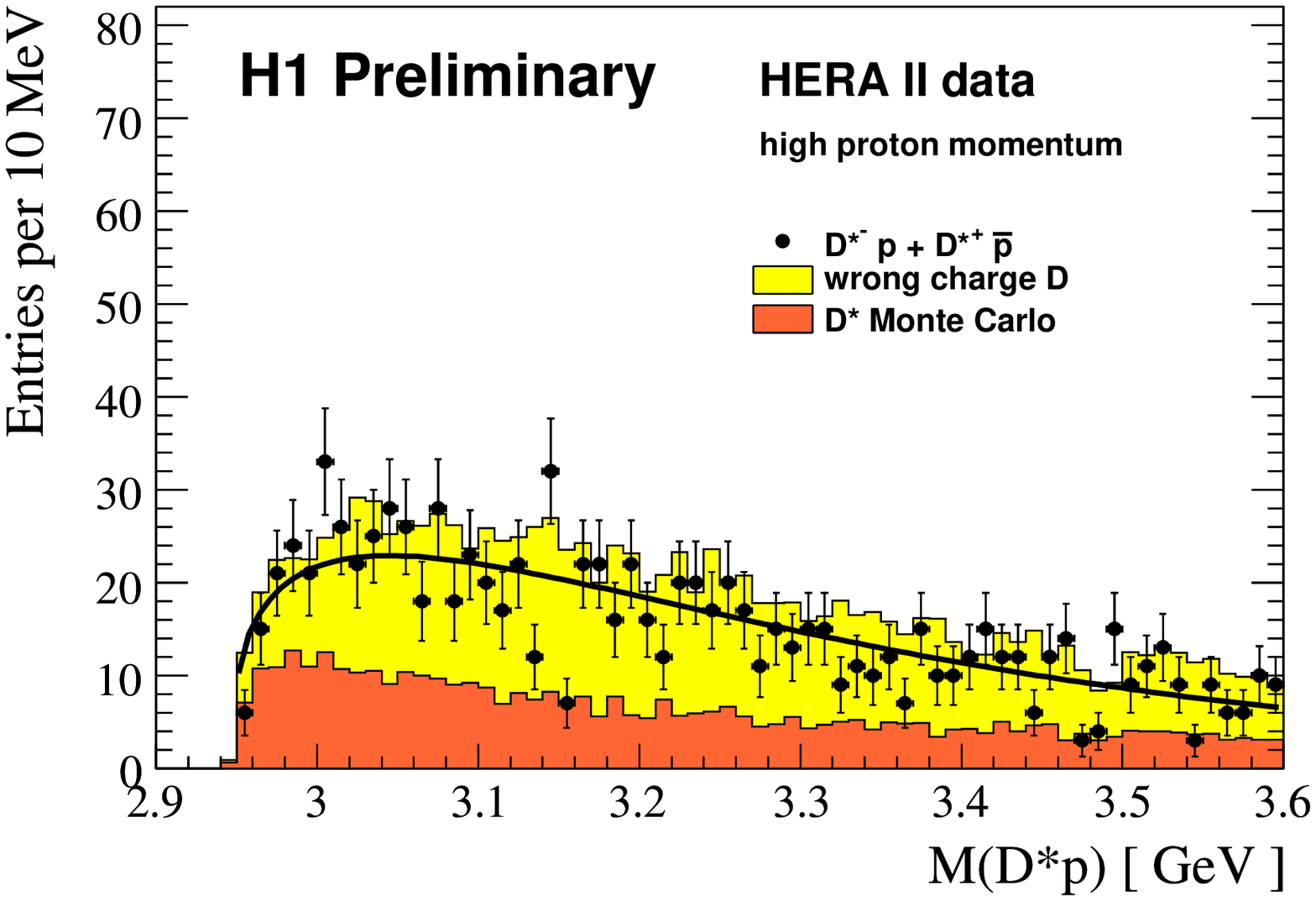}
\includegraphics[bb = 0 0 550 390, clip=true,width=0.38\textwidth]{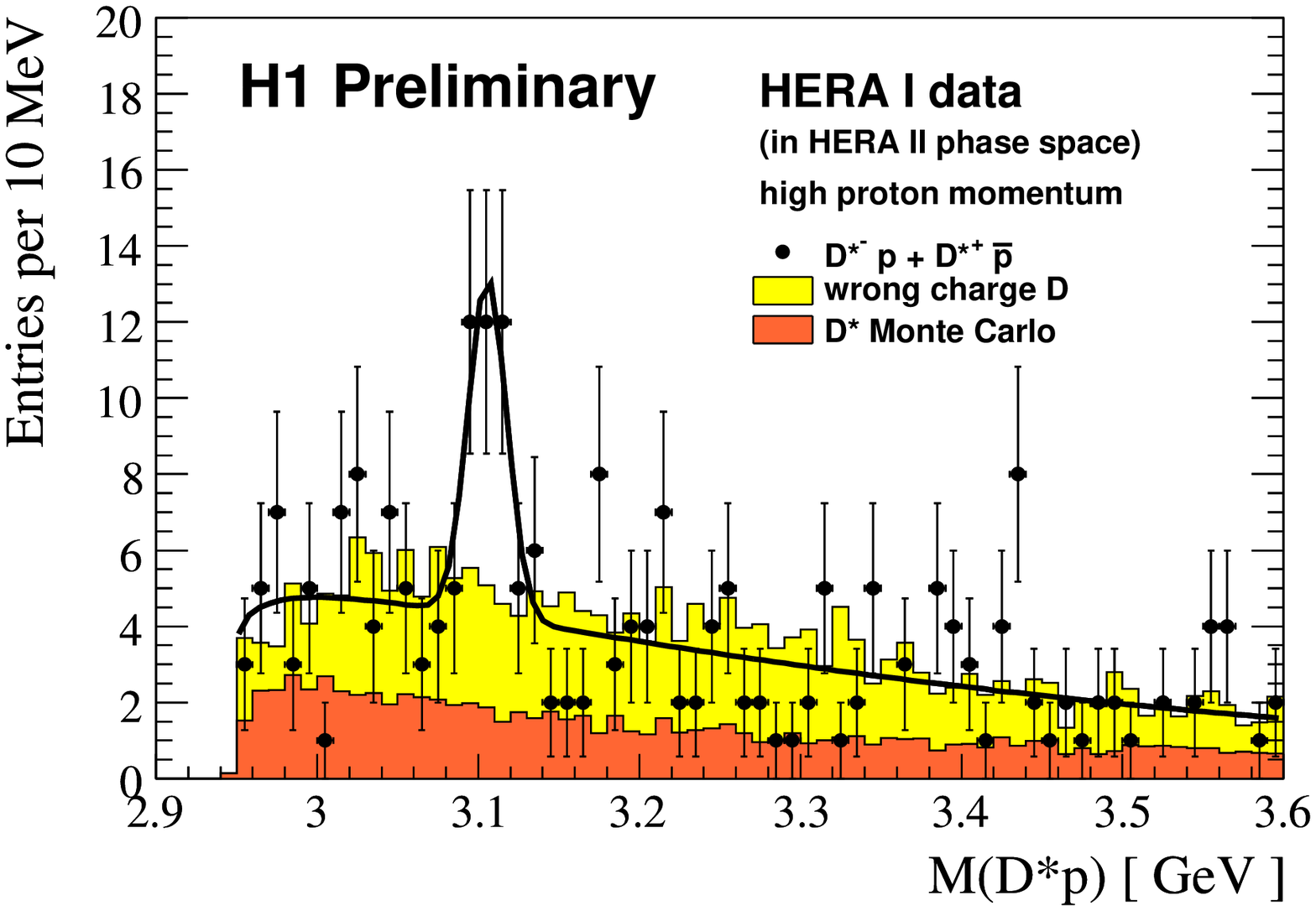}
\end{center}
  \vspace{-20pt}
\caption{Distribution of $M(D^{*}p$ in the HERA\,I (bottom) and
HERA\,II (top) data samples.}
\label{pent}
\end{wrapfigure}
The H1 collaboration observed a narrow signal in the $D^{*}p$ channel at $M(D^{*}p)= 3099\pm 3(\text{stat.})\pm 5(\text{syst.})$
MeV in the HERA\,I dataset~\cite{h1pent1:08}.
This signal was interpreted as an anti-charm baryon with a minimal constituent
quark composition of $uudd\bar{c}$ with a relative contribution of
$\frac{N(D^{*}p)}{N(D^{*})}\sim 1 \%$ in the identified $D^{*}$ meson sample.
In contrast to this, no evidence was found by the ZEUS~\cite{zeuspent:08} collaboration.
Using the full HERA dataset H1 does not confirm the observation.
The data of 2004 to 2007 correspond to an integrated luminosity of
$\unit{348}{\,\text{pb}^{-1}}$ and thus increase the available statistics by a
factor of $\sim 4$.
In contrast to the previous analysis of the $D^{*}p$ final state no dE/dx
requirements were used for particle identification. A
detailed description of the event selection can be found in~\cite{h1pent2:08}.
The mass spectrum was reconstructed both for the HERA\,II dataset as well as for
the HERA\,I dataset using the new selection criteria.
Figure~\ref{pent} shows the reconstructed invariant mass for the $D^{*}p$
candidates separately for the two datasets. In the reanalysed HERA\,I dataset the previously
reported signal is reproduced. Under the assumption of the width from the
HERA\,I measurement there is no excess in the HERA\,II data visible.
The upper limit at a 95 \% confidence limit was
calculated to be $\frac{N(D^{*}p)}{N(D^{*})}\sim 0.1 \%$.

\end{document}